\documentclass[aps, prb, twocolumn, superscriptaddress]{revtex4-2}
\usepackage{amsmath, amsfonts, amssymb, ascmac, mathtools, bm, braket, comment}
\usepackage{multirow}
\usepackage{graphicx}
\usepackage{float, color, xcolor}

\usepackage{bbm}
\usepackage{tabularx}

\usepackage{tikz}
\usetikzlibrary{shapes.geometric}

\usepackage[
pagebackref=false,
colorlinks=true,
linkcolor=blue,
urlcolor=blue,
filecolor=black,
citecolor=red,
pdfstartview=FitV,
pdftitle={},
pdfauthor={},
pdfsubject={},
pdfkeywords={},
pdfpagemode=None,
bookmarksopen=true
]{hyperref}

\newcommand{\ii}{\text{i}}
\newcommand{\tr}{\mathrm{tr}}
\newcommand{\RL}{\mathrm{RL}}

\DeclareRobustCommand{\boxmarker}{
  \tikz[baseline=-0.5ex]{
    \node[
      draw=blue, fill=white, thick, rectangle,
      inner sep=0pt,
      minimum width=.5em, minimum height=.5em
    ] {};
  }
}
\DeclareRobustCommand{\trianglemarker}{
  \tikz[baseline=-0.5ex]{
    \node[
      isosceles triangle,
      minimum width=0.2cm, minimum height=0.2cm,
      thick,
      rotate=90,
      isosceles triangle apex angle=60,
      draw=purple, fill=white,
      inner sep=0pt
    ] {};
  }
}
\DeclareRobustCommand{\circlemarker}{%
  \tikz[baseline=-0.5ex]{
    \node[
      draw=teal, fill=white, thick, circle, minimum size=0.2cm,
      inner sep=0em
    ] {};
  }
}

\begin{document}

\title{Complex entanglement entropy for complex conformal field theory}

\author{Haruki Shimizu}
\email{shimizu-haruki@issp.u-tokyo.ac.jp}
\affiliation{Institute for Solid State Physics, University of Tokyo, Kashiwa, Chiba 277-8581, Japan}

\author{Kohei Kawabata}
\email{kawabata@issp.u-tokyo.ac.jp}
\affiliation{Institute for Solid State Physics, University of Tokyo, Kashiwa, Chiba 277-8581, Japan}

\date{\today}

\begin{abstract}
Conformal field theory underlies critical ground states of quantum many-body systems.
While conventional conformal field theory is associated with positive central charges, nonunitary conformal field theory with complex-valued central charges has recently been recognized as physically relevant.
Here, we demonstrate that complex-valued entanglement entropy characterizes complex conformal field theory and critical phenomena of open quantum many-body systems.
This is based on non-Hermitian reduced density matrices constructed from the combination of right and left ground states.
Applying the density matrix renormalization group to non-Hermitian systems, we numerically calculate the complex entanglement entropy of the non-Hermitian five-state Potts model, thereby confirming the scaling behavior predicted by complex conformal field theory.
\end{abstract}

\maketitle

\section{Introduction}

Quantum entanglement is a cornerstone of quantum physics~\cite{Nielsen-textbook}.
The spread of entanglement lays the foundations for quantum chaos and thermalization~\cite{Calabrese-Cardy-05, *Calabrese-Cardy-06, Kim-13, Nahum-17}.
Entanglement also provides a framework for capturing phase transitions and critical phenomena in quantum many-body systems~\cite{Osterloh-02, Osborne-02}.
Specifically, conformal invariance emerges at critical points, leading to the description in terms of conformal field theory (CFT).
In $1+1$ dimensions, this results in the universal scaling behavior of von Neumann entanglement entropy~\cite{Holzhey-94, Vidal-03, Calabrese-Cardy-04, Calabrese-Cardy-review},
\begin{equation}
    S = \frac{c}{3} \log l,
        \label{eq: EE CFT}
\end{equation}
where $l$ denotes the subsystem length.
The central charge $c$, a positive number for conventional unitary CFT, contains the general characteristics of the universality class.
Entanglement also plays a key role in other types of quantum phases, such as topological phases of matter~\cite{Kitaev-Preskill-06, Levin-Wen-06, Ryu-06, Li-Haldane-08, Pollmann-10}.
Moreover, it characterizes nonunitary quantum phase transitions induced by the interplay between unitary dynamics and measurements~\cite{Chan-19, Skinner-19, Li-18, *Li-19, Choi-20, Gullans-20, Jian-20, Fisher-review}.

The Potts model is a paradigmatic model of phase transitions, representing a $Q$-state generalization of the Ising model~\cite{Potts-52, Wu-82}.
Despite its apparent simplicity, it exhibits rich statistical-mechanical behavior.
While it undergoes a phase transition for any value of $Q$, the nature of this phase transition depends on $Q$.
Extensive analytical and numerical studies established that the phase transition in $1+1$ dimensions is continuous for $Q \leq 4$ but becomes discontinuous for 
$Q > 4$~\cite{Baxter-73, Nienhuis-79, Nienhuis-80, Nauenberg-80, Binder-81, Cardy-80, Nienhuis-81, Dotsenko-84, Dotsenko-Fateev-84, Zamolodchikov-85, Zamolodchikov-87, Klumper-89, Buddenoir-93, Nishino-98, Delfino-00, Ozeki-03, Shun-14, Liang-17, Iino-19, Morita-19, DEmidio-23}.
In particular, for $Q = 5$, the discontinuous phase transition nearly resembles a continuous phase transition due to a small energy gap and a large correlation length.
As a result, approximate conformal invariance appears even in the absence of genuine quantum criticality.

Notably, the approximate conformal invariance in the five-state Potts model originates from complex CFT~\cite{Kaplan-09, Wang-17, Gorbenko-18a, *Gorbenko-18b, Ma-19, Benini-20, Faedo-20, Giombi-20, Nahum-22, Han-23, Haldar-23, Jacobsen-24, Tang-24}.
Once we complexify the coupling parameter of the Potts model with $Q > 4$ through analytic continuation, a pair of fixed points emerges for the complex-valued coupling parameters.
These complex fixed points are described by CFT with the complex-valued central charges $c \in \mathbb{C}$, as opposed to unitary CFT possessing positive central charges $c > 0$.
Complex CFT is distinguished from a different type of nonunitary CFT characterized by negative central charges $c < 0$, exemplified by the Yang-Lee edge singularity~\cite{Yang-52, *Lee-52, Fisher-78, Cardy-85} and other related non-Hermitian models~\cite{Couvreur-17, Chang-20, Lee-22, Guo-21, Tu-22, Ryu-23, Hsieh-23, Fossati-23, Rottoli-24, Xue-24} involving spontaneous breaking of parity-time symmetry~\cite{Bender-02, Bender-review}.
It should also be noted that such non-Hermitian models can be physically realized in open quantum systems, where exchanges of energy and particles with the external environment effectively induce non-Hermiticity~\cite{Konotop-review, Christodoulides-review}.

Remarkably, recent studies constructed microscopic lattice models corresponding to complex CFT~\cite{Jacobsen-24, Tang-24}.
As a direct consequence of complex CFT, the resulting lattice models no longer respect Hermiticity.
Numerical calculations on the finite-size scaling of complex eigenenergies demonstrated the complex central charges and complex scaling dimensions~\cite{Tang-24}.
However, the scaling behavior of entanglement entropy in Eq.~(\ref{eq: EE CFT}) has yet to be verified for this non-Hermitian Potts model.
If we directly apply Eq.~(\ref{eq: EE CFT}) to complex CFT with $c \in \mathbb{C}$, the resulting entanglement entropy also becomes complex valued, which conflicts with the conventional understanding that entanglement entropy is a real-valued quantity.
Thus, the role of quantum entanglement in complex CFT, as well as in open quantum many-body systems, has remained elusive.

In this work, we demonstrate that the universal scaling behavior of entanglement entropy in Eq.~(\ref{eq: EE CFT}) is still applicable even to complex CFT with complex central charges $c \in \mathbb{C}$.
We introduce complex-valued entanglement entropy based on non-Hermitian reduced density matrices constructed from the combination of right and left eigenstates.
While such generalized entanglement entropy was employed to characterize nonunitary CFT with negative central charges~\cite{Couvreur-17, Chang-20, Lee-22, Guo-21, Tu-22, Ryu-23, Hsieh-23, Fossati-23, Rottoli-24, Xue-24} and chaotic behavior of open quantum systems~\cite{Cipolloni-23}, it has never been applied to complex CFT.
It is also noteworthy that complex entanglement entropy is referred to as pseudo-entropy in the recent literature of high energy physics~\cite{Nakata-21, Mollabashi-21, *Mollabashi-21R, Murciano-22, Guo-23, Doi-23, *Doi-23JHEP, Narayan-23, Kanda-23, Kanda-24}.
Our results show the utility of complex entanglement entropy in non-Hermitian quantum many-body systems.
Using the density matrix renormalization group (DMRG), we numerically calculate the complex entanglement entropy of the non-Hermitian five-state Potts model.
We confirm the agreement with the underlying complex CFT, thereby validating the significant role of entanglement entropy in open quantum many-body systems.

The rest of this work is organized as follows.
In Sec.~\ref{sec: model}, we introduce the non-Hermitian Potts model~\cite{Tang-24}.
In Sec.~\ref{sec: pseudo entropy}, we describe the construction of complex entanglement entropy.
In Sec.~\ref{sec: DMRG}, we present our numerical results of complex entanglement entropy based on DMRG.
In Sec.~\ref{sec: discussion}, we conclude this work.

\section{Non-Hermitian Potts model}
    \label{sec: model}

We begin with the Hermitian $Q$-state quantum Potts model in $1+1$ dimensions ($J, h \in \mathbb{R}$),
\begin{equation}
    \hat{H}_0 = - \sum_{i=1}^{L} \sum_{k=1}^{Q-1} \left[ J\,( \hat{\sigma}_i^{\dag} \hat{\sigma}_{i+1})^{k} + h \hat{\tau}_i^k \right],
        \label{eq: Potts Hermitian}
\end{equation}
where $\hat{\sigma}_i$ denotes the spin phase operator at site $i$, and $\hat{\tau}_i$ the spin shift operator. 
We impose $\hat{\sigma}_{L+1}\coloneqq\hat{\sigma}_1,\hat{\tau}_{L+1}\coloneqq\hat{\tau}_1$ for periodic boundary conditions and $\hat{\sigma}_{L+1}\coloneqq 0,\hat{\tau}_{L+1}\coloneqq 0$ for open boundary conditions.
These operators act on the local degrees of freedom at each site, represented by $\{ \ket{n} | n=1, 2, \cdots, Q\}$, and are defined to satisfy the relations,
\begin{align}
    \hat{\sigma} \ket{n} &= e^{2\pi\ii n/Q} \ket{n}, \\
    \hat{\tau} \ket{n} &= \ket{n+1 \left( \mathrm{mod}~Q \right)},
\end{align}
with their matrix representations given by
\begin{align}
    \hat{\sigma} &= \begin{pmatrix}
        e^{2\pi\ii/Q} \\
        & e^{4\pi\ii/Q} \\
        & & \ddots \\
        & & & e^{2 \left( Q-1\right)\pi\ii/Q} \\
        & & & & 1
    \end{pmatrix}, \\
    \hat{\tau} &= \begin{pmatrix}
        & & & & 1\\
        1 & \\
        & 1 & \\
        & & \ddots \\
        & & & 1 &
    \end{pmatrix}.
\end{align}
For $Q=2$, the Potts model reduces to the transverse-field Ising model.
The Hamiltonian $\hat{H}_0$ is symmetric under spin permutation and undergoes a quantum phase transition at the point of spontaneous symmetry breaking.
The phase transition occurs precisely at $J = h$, as determined by the Kramers-Wannier duality.
Extensive analytical and numerical studies established that this phase transition is continuous for $Q \leq 4$ but becomes discontinuous for $Q > 4$~\cite{Baxter-73, Nienhuis-79, Nienhuis-80, Nauenberg-80, Binder-81, Cardy-80, Nienhuis-81, Dotsenko-84, Dotsenko-Fateev-84, Zamolodchikov-85, Zamolodchikov-87, Klumper-89, Buddenoir-93, Nishino-98, Delfino-00, Ozeki-03, Shun-14, Liang-17, Iino-19, Morita-19, DEmidio-23}.

For $Q=5$, the discontinuous phase transition is exceptionally weak and almost continuous because of a small energy gap and a large correlation length.
This weakly discontinuous nature is attributed to the presence of fixed points with complexified coupling parameters.
These complex fixed points accompany conformal invariance characterized by a complex-valued central charge~\cite{Kaplan-09, Wang-17, Gorbenko-18a, *Gorbenko-18b, Ma-19, Benini-20, Faedo-20, Giombi-20, Nahum-22, Han-23, Haldar-23, Jacobsen-24, Tang-24}.

Recent works constructed  microscopic lattice models corresponding to complex CFT~\cite{Jacobsen-24, Tang-24}.
In this work, we employ the non-Hermitian Potts model $\hat{H}_0 + \hat{H}_1$ in Ref.~\cite{Tang-24}, where $\hat{H}_0$ is the Hermitian Potts model in Eq.~(\ref{eq: Potts Hermitian}), and the non-Hermitian perturbation $\hat{H}_1$ reads
\begin{align}
    \hat{H}_1 &= \lambda \sum_{i=1}^{L} \sum_{k_1, k_2 = 1}^{Q-1} \left[ (\hat{\tau}_i^{k_1} + \hat{\tau}_{i+1}^{k_1})\,(\hat{\sigma}_i^{\dag}\hat{\sigma}_{i+1})^{k_2} \right. \nonumber \\
    &\qquad\qquad\qquad\qquad \left. + (\hat{\sigma}_i^{\dag}\hat{\sigma}_{i+1})^{k_1}\,(\hat{\tau}_i^{k_2} + \hat{\tau}_{i+1}^{k_2}) \right]
\end{align}
with a complex-valued parameter $\lambda \in \mathbb{C}$.
Despite non-Hermiticity, the perturbation $\hat{H}_1$ also preserves spin permutation symmetry and Kramers-Wannier duality.
Reference~\cite{Tang-24} numerically identified the complex fixed points for $Q=5$ as
\begin{equation}
    J_{\rm c} = h_{\rm c} = 1, \quad \lambda_{\rm c} = 0.079 \pm 0.060\ii.
        \label{eq: lambda-c}
\end{equation}
In the following, we fix the parameters as these critical values.
The CFT analysis based on the analytic continuation to the complex coupling predicts the complex central charge~\cite{Gorbenko-18b}
\begin{align}
    c &= 1 + \frac{3}{\pi} \frac{\left[ \cosh^{-1} \left( Q/2-1\right) \right]^2}{2\pi \mp \ii \cosh^{-1} \left( Q/2-1 \right)} \nonumber \\
    &\approx 1.13755 \pm 0.0210687\ii \quad \left( Q=5 \right).
        \label{eq: c-CFT}
\end{align}
On the other hand, Ref.~\cite{Tang-24} numerically obtained
\begin{equation}
    c \approx 1.1405 \pm 0.0224\ii
        \label{eq: c-Tang-24}
\end{equation}
from the finite-size scaling of complex eigenenergies.
In this work, we numerically determine the complex central charge through the finite-size scaling of complex entanglement entropy.

\section{Complex entanglement entropy}
    \label{sec: pseudo entropy}

Before introducing complex entanglement entropy, we first provide a review of the conventional framework for calculating entanglement entropy.
We divide a quantum system (or Hilbert space) into two parts, A and B.
For a given quantum state $\ket{\psi}$ in the entire system, the corresponding density matrix reads
\begin{equation}
    \hat{\rho} = \ket{\psi} \bra{\psi}.
        \label{eq: density matrix}
\end{equation}
We introduce the reduced density matrix in subsystem A by
\begin{equation}
    \hat{\rho}_{\rm A} \coloneqq \tr_{\rm B}\,\hat{\rho},
        \label{eq: reduced density matrix}
\end{equation}
where the trace is taken over the Hilbert space restricted to subsystem B.
The von Neumann entanglement entropy is then given as
\begin{equation}
    S \coloneqq - \tr_{\rm A} \left( \hat{\rho}_{\rm A} \log \hat{\rho}_{\rm A} \right).
        \label{eq: vN EE def}
\end{equation}
For a critical ground state of a Hermitian system in $1+1$ dimensions, this leads to the CFT scaling behavior in Eq.~(\ref{eq: EE CFT}).
Even in non-Hermitian Hamiltonians, we can calculate entanglement entropy for a given eigenstate by Eq.~(\ref{eq: vN EE def}).
The entanglement entropy obtained in this manner remains real valued owing to Hermiticity of the density matrix $\hat{\rho}$ in Eq.~(\ref{eq: density matrix}).
Consequently, this real-valued entropy cannot reproduce the CFT scaling behavior in Eq.~(\ref{eq: EE CFT}) for complex-valued central charges $c \in \mathbb{C}$.

To capture complex CFT, we note that a hallmark of non-Hermitian Hamiltonians is the distinction between right and left eigenstates~\cite{Brody-14}.
For a given non-Hermitian Hamiltonian $\hat{H}$ and its eigenenergy $E \in \mathbb{C}$, the eigenequation reads
\begin{equation}
    \hat{H} \ket{\psi} = E \ket{\psi}, \quad \hat{H}^{\dag} | \psi \rangle\!\rangle = E^{*} | \psi \rangle\!\rangle.
\end{equation}
Here, $\ket{\psi}$ and $| \psi \rangle\!\rangle$ are right and left eigenstates, respectively, which are not necessarily identical, unlike in the Hermitian case.
This observation motivates the introduction of  a generalized density matrix and its reduced density matrix~\cite{Chang-20},
\begin{equation}
    \hat{\rho}^{\RL} \coloneqq \frac{\ket{\psi} \langle\!\langle \psi|}{\langle\!\langle \psi \ket{\psi}}, \quad
    \hat{\rho}^{\RL}_{\rm A} \coloneqq \tr_{\rm B}\,\hat{\rho}^{\RL}.
        \label{eq: pseudo reduced density matrix}
\end{equation}
In contrast with the conventional density matrix in Eq.~(\ref{eq: density matrix}), $\hat{\rho}^{\RL}$ no longer respects Hermiticity:
\begin{equation}
    (\hat{\rho}^{\RL})^{\dag} \neq \hat{\rho}^{\RL}.
\end{equation}
As a result, its eigenvalues are generally complex valued, and hence the generalized von Neumann entropy, 
\begin{equation}
    S \coloneqq - \tr_{\rm A}\,( \hat{\rho}^{\RL}_{\rm A} \log \hat{\rho}^{\RL}_{\rm A} ),
        \label{eq: pseudo entropy}
\end{equation}
is also complex.
In a similar manner, the generalized $n$th R\'enyi entropy is defined by
\begin{equation}
    S_n \coloneqq \frac{1}{1-n} \log \tr_{\rm A}\,( (\hat{\rho}^{\RL}_{\rm A})^n ),
\end{equation}
which reduces to the generalized von Neumann entropy $S$ in Eq.~(\ref{eq: pseudo entropy}) for $n \to 1$.
While Eq.~(\ref{eq: pseudo entropy}) depends on the branch choice of the complex logarithm, we consistently adopt the following convention throughout this work:
\begin{equation}
    \log z = \log \left| z \right| + \ii \operatorname{arg} z, \quad \operatorname{arg} z \in (-\pi, \pi]
\end{equation}
for $z \in \mathbb{C}$.

In Refs.~\cite{Chang-20, Lee-22, Guo-21, Xue-24}, the generalized entanglement entropy in Eq.~(\ref{eq: pseudo entropy})
was employed to demonstrate nonunitary CFT with negative central charges $c < 0$.
For this class of nonunitary CFT, the entanglement entropy remains real.
By contrast, here we show that the generalized entanglement entropy also captures nonunitary CFT even with complex central charges $c \in \mathbb{C}$.
It is also noteworthy that Refs.~\cite{Couvreur-17, Tu-22, Hsieh-23, Fossati-23} adopted an alternative definition of the generalized entanglement entropy,
\begin{equation}
    S_{\mathrm{abs}} \coloneqq - \sum_{i} \lambda_i \log \left| \lambda_i \right|, \label{eq: abs-entropy}
\end{equation}
where $\lambda_i \in \mathbb{C}$ denotes the complex eigenvalues of the non-Hermitian reduced density matrix $\hat{\rho}^{\RL}_{\rm A}$.
This definition is free from the ambiguity associated with the logarithm, in contrast to Eq.~(\ref{eq: pseudo entropy}).
In Sec.~\ref{subsec: other}, we also present numerical calculations of $S_{\mathrm{abs}}$ for the critical non-Hermitian Potts model.

The generalized reduced density matrix in Eq.~(\ref{eq: pseudo reduced density matrix}) inherits symmetry and other relevant properties of the original non-Hermitian Hamiltonian.
In the absence of spectral degeneracy, the non-Hermitian Hamiltonian is generally diagonalized as
\begin{equation}
    \hat{H} = \sum_{n} E_n \frac{\ket{\psi_n} \langle\!\langle \psi_n |}{\langle\!\langle \psi_n \ket{\psi_n}},
\end{equation}
where $E_n$ represents a complex eigenenergy, and $\ket{\psi_n}$ ($| \psi_n \rangle\!\rangle$) denotes the corresponding right (left) eigenstate.
Thus, $\ket{\psi_n} \langle\!\langle \psi_n |/\langle\!\langle \psi_n \ket{\psi_n}$ serves as a non-Hermitian projector to the eigenspace indexed by $n$.
It is also notable that the definition of the non-Hermitian reduced density matrix $\hat{\rho}^{\RL}$ in Eq.~(\ref{eq: pseudo reduced density matrix}) requires $\langle\!\langle \psi \ket{\psi} \neq 0$.
This condition is satisfied even in non-Hermitian Hamiltonians when there is no degeneracy in the complex spectrum~\cite{Brody-14}.
However, this condition can be violated (i.e., $\langle\!\langle \psi \ket{\psi} = 0$) if the eigenstates $\ket{\psi}$ experience spectral degeneracy, which is a defining feature of exceptional points~\cite{Kato-textbook, Berry-04, Heiss-12}.
Such an ill-defined nature can be the origin of the universal scaling behavior in Eq.~(\ref{eq: EE CFT}) for nonunitary CFT.
However, in actual numerical calculations of finite-size systems, the spectral degeneracy
between the ground and first excited states 
is typically lifted with small level spacing due to the finite-size effect, as also observed in our subsequent numerical calculations (see Sec.~\ref{subsec: Petermann} for further discussion).

\section{Complex conformal field theory}
    \label{sec: DMRG}

\subsection{Density matrix renormalization group}

To obtain the right ground state $\ket{\psi}$ and left ground state $|\psi\rangle\!\rangle$ of the critical non-Hermitian five-state Potts model $\hat{H} = \hat{H}_0 + \hat{H}_1$, we employ two-site DMRG for finite systems and utilize the matrix product state (MPS) representation of wave functions.
While the spectrum of $\hat{H}$ is generally complex valued, here we choose its ground state as the eigenstate with the minimum real part of complex eigenenergy, compatible with the CFT analysis.

DMRG mainly consists of an iteration of two procedures,
(A)~diagonalization of the projected Hamiltonian acting on two sites and
(B)~selection of the projection operator to construct the projected Hamiltonian shifted by one site in preparation for the next step.
For Hermitian Hamiltonians, the Lanczos method is usually adopted in step (A).
In step~(B), one takes the largest $D$ singular values of $\ket{\psi}$ divided at the middle of the two sites to construct an optimal MPS with bond dimension $D$, whose optimality is explained as follows~\cite{Carlon1999}.

Let $\ket{\psi} = \sum_{i, j} \psi_{ij}\ket{i}\ket{j}$ ($\psi_{ij} \in \mathbb{C}$) be a target state divided into left and right subsystems, where $\{ \ket{i} \}$ and $\{ \ket{j} \}$ denote certain bases in the left and right subsystems, respectively.
This wave function is approximated by
\begin{equation}
    \ket{\tilde{\psi}} = \sum_{i,j,\alpha} u_{i\alpha}\tilde{\psi}_{\alpha j}\ket{i}\ket{j},
\end{equation}
where $\alpha \in \{1,\ldots,D\}$ is the index of a restricted basis to approximate the original basis, 
and $u_{i\alpha} \in \mathbb{C}$ is the isometry satisfying $\sum_{i} u_{i\alpha}^\ast u_{i\beta} = \delta_{\alpha\beta}$.
The accuracy of the approximation is evaluated by the norm of the difference:
\begin{align}
    & \lVert \ket{\psi} - \ket{\tilde{\psi}} \rVert^2 \notag
    \\  
    &\quad = 1 - \sum_{i,j,\alpha} \left[\psi^\ast_{ij} (u_{i\alpha}\tilde\psi_{\alpha j}) + \mathrm{c.c.}\right] + \sum_{\alpha, j}\tilde\psi_{\alpha j}^\ast\tilde\psi_{\alpha j}.
        \label{eq: psi-diff}
\end{align}
The global minimum point of $\lVert \ket{\psi} - \ket{\tilde{\psi}} \rVert^2$ can be found using the method of Lagrange multipliers.
Under the condition of $\sum_{i} u_{i\alpha}^\ast u_{i\beta} = \delta_{\alpha\beta}$, the Lagrangian is
\begin{equation}
    L = \lVert \ket{\psi} - \ket{\tilde{\psi}} \rVert^2 + \sum_{\alpha,\beta}\lambda_{\alpha\beta}\left(\sum_i u_{i\alpha}^\ast u_{i\beta} - \delta_{\alpha\beta}\right),
\end{equation}
where each $\lambda_{\alpha\beta}$ is the Lagrange multiplier. 
Without loss of generality, we can assume $\lambda$ to be a Hermitian matrix: $\lambda_{\alpha\beta} = \lambda^\ast_{\beta\alpha}$. 
Making $L$ stationary with respect to $u_{i\alpha},u^\ast_{i\alpha},\tilde\psi_{\alpha j},\tilde\psi_{\alpha j}^\ast$, and $\lambda_{\alpha\beta}$, we have
\begin{align}
    \tilde\psi_{\alpha j} &= \sum_i u_{i\alpha}^\ast \psi_{ij}, \label{eq: tpsi-upsi}\\
    \sum_j \psi^\ast_{ij} \tilde\psi_{\beta j} &= \sum_\alpha u^\ast_{i\alpha}\lambda_{\alpha\beta}. 
        \label{eq: psitpsi-ulambda}
\end{align}
From Eqs.~\eqref{eq: tpsi-upsi} and \eqref{eq: psitpsi-ulambda}, we obtain the relationship between the isometry $u_{i\alpha}$ and the density matrix $\rho_{ik}\coloneqq \sum_j \psi^\ast_{ij}\psi_{kj}$,
\begin{equation}
    \sum_k \rho_{ik} u^\ast_{k\beta} = \sum_\alpha u^\ast_{i\alpha} \lambda_{\alpha\beta}.
\end{equation}
Diagonalizing $\lambda$ as $\lambda_{\alpha\beta} = \sum_\gamma v^\ast_{\alpha\gamma}\lambda_\gamma v_{\beta\gamma}$ with a unitary matrix $v$, we further have
\begin{equation}
    \sum_{k,\beta} \rho_{ik} u^\ast_{k\beta}v^\ast_{\beta\gamma} = \sum_{\alpha,\gamma} u^\ast_{i\alpha} v^\ast_{\alpha\gamma} \lambda_\gamma,
\end{equation}
which implies that each $\lambda_\gamma$ is the eigenvalue of $\rho$. 
Thus, at the stationary point, Eq.~(\ref{eq: psi-diff}) reduces to
\begin{align}
    \lVert \ket{\psi} - \ket{\tilde{\psi}} \rVert^2 &= 1 - \sum_{i,k,\alpha} u_{i\alpha} \rho_{ik} u^\ast_{k\alpha} = 1 - \sum_\gamma \lambda_\gamma. \label{eq: one-sumlambda}
\end{align}
From Eq.~\eqref{eq: one-sumlambda}, the best approximation is given by letting $u$ take $D$ eigenvectors associated with the largest $D$ eigenvalues of $\rho$. 
Since the eigenvalues of $\rho$ are the square of the singular values of $\psi$, we can implement the singular value decomposition (SVD) of $\psi$ instead of the eigenvalue diagonalization of $\rho$.

For non-Hermitian Hamiltonians, there are several possible options for both (A) and (B)~\cite{Carlon1999, Yamamoto-22}.
Here, we obtain the right and left eigenstates one by one.
We adopt the Arnoldi method to obtain the right eigenstate of the projected Hamiltonian for step~(A) and minimize $\| \ket{\psi} - \ket{\tilde\psi} \|$ using SVD for step~(B).
Then, we obtain the corresponding left eigenstate $|\psi\rangle\!\rangle$ of the Hamiltonian $\hat{H}$ of eigenenergy $E$ as the right eigenstate of $\hat{H}^\dagger$ of eigenenergy $E^\ast$.
In all of our numerical calculations, we use the \textsc{ITensor} library~\cite{10.21468/SciPostPhysCodeb.4, 10.21468/SciPostPhysCodeb.4-r0.3}.
In our five-state non-Hermitian Potts model,
since the Hamiltonian $\hat{H}$ is symmetric (i.e., $\hat{H}^{T} = \hat{H}$), left eigenstates $|\psi\rangle\!\rangle$ are merely complex conjugates of the corresponding right eigenstates $\ket{\psi}$ (i.e., $|\psi\rangle\!\rangle \propto \ket{\psi}^{*}$).
While we do not explicitly employ this property in our actual numerical calculations, we confirm that the numerical difference $\| |\psi\rangle\!\rangle - \ket{\psi}^{*} \|$ is around $3 \times 10^{-5}$ for $L=24$ and $D=400$.
Furthermore, while here we take the simplest way, it is also possible to improve accuracy by adopting $\hat{\rho}_{\rm A} = \tr_{\rm B}\,(\ket{\psi}\bra{\psi} + \ket{\psi}\!\rangle\langle\!\bra{\psi})/2$ as the density matrix used for truncation~\cite{Carlon1999,Yamamoto-22}, with all of their matrix elements being real as long as the Hamiltonian is symmetric.

\subsection{Periodic boundary conditions}

\begin{figure}[tb]
\centering
\includegraphics[width=\linewidth]{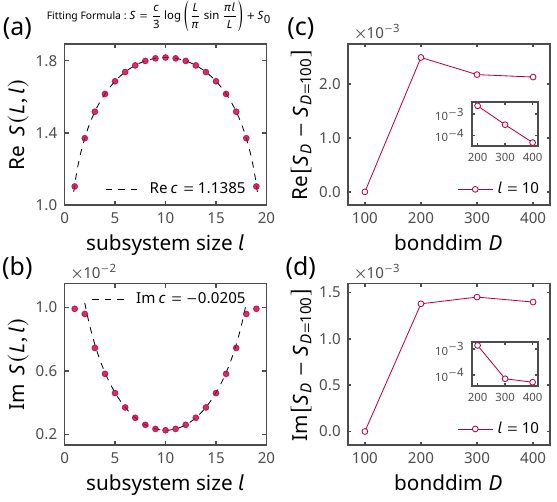}
\caption{(a) Real and (b) imaginary parts of complex entanglement entropy as functions of the subsystem size $l$, numerically obtained through the density matrix renormalization group for the system size $L=20$ and bond dimension $D=400$ under the periodic boundary conditions.
We fit the numerical data using the three points in the center.
(c)~Real and (d) imaginary parts of complex entanglement entropy compared to that of $D=100$ [i.e., $S_{D} \left( L, L/2 \right) -S_{D=100} \left( L, L/2 \right)$] with respect to the bond dimension $D$.
Insets: The changes in entropy compared to the $100$ smaller bond dimension [i.e., $S_D \left( L, L/2 \right) - S_{D-100}\left( L, L/2 \right)$] for each $D = 200, 300, 400$.}
    \label{fig: complex-entropy-pbc}
\end{figure}

We numerically obtain the complex entanglement entropy in Eq.~\eqref{eq: pseudo entropy} for the ground state of the critical non-Hermitian five-state Potts model $\hat{H}$ with periodic boundaries. The numerical results shown in Figs.~\ref{fig: complex-entropy-pbc}\,(a) and \ref{fig: complex-entropy-pbc}\,(b) agree well with the finite-size scaling of CFT~\cite{Calabrese-Cardy-04, Calabrese-Cardy-review},
\begin{equation}
    S^{\mathrm{PBC}} (L, l) = \frac{c}{3} \log \left( \frac{L}{\pi} \sin \frac{\pi l}{L} \right) + S_0,
\end{equation}
especially around the center $l \simeq L/2$.
Figures~\ref{fig: complex-entropy-pbc}\,(c) and \ref{fig: complex-entropy-pbc}\,(d) show the change in the entanglement entropy with respect to the bond dimension $D$, suggesting that the further increase in $D$ more than $400$ should not cause a change larger than $10^{-3}$ in the entanglement entropy and thus in the obtained central charge.

\begingroup
\setlength{\tabcolsep}{9pt} 
\renewcommand{\arraystretch}{1.2} 
\begin{table}[tb]
  \centering
  \caption{Complex central charge $c$.
  The row ``CFT" shows $c$ derived from the complex conformal field theory (CFT) [i.e., Eq.~(\ref{eq: c-CFT})].
  The rows ``$L=4$", $\cdots$, ``$L=24$" show $c$ numerically obtained from the fitting of complex entanglement entropy for the non-Hermitian reduced density matrices $\hat{\rho}_{\rm A}^{\rm RL}$.
  The row ``$L=24$ ($\hat{\rho}_{\rm A}^{\rm R}$)" shows $c$ for the Hermitian reduced density matrices $\hat{\rho}_{\rm A}^{\rm R}$ constructed solely from the right ground state.
  The row ``$L=24$ ($\hat{\rho}_{\rm A}^{\rm SVD}$)" shows $c$ of singular value decomposition (SVD) entanglement entropy for $\hat{\rho}_{\rm A}^{\rm SVD}$.
  The numerical results are based on the density matrix renormalization group for bond dimensions $D=300$ and $D=400$.}
  \label{tab: central-charges}
  \begin{tabular}{lll}
  \hline \hline
  & $c\, (D=300)$ & $c\, (D=400)$\\ \hline 
  CFT  & & $1.1376 - 0.0211 \ii$ \\ \hline
  $L=4$ & $1.1927 + 0.0104 \ii$ & $1.1927 + 0.0104 \ii$ \\
  $L=8$ & $1.1429 - 0.0171\ii$ & $1.1429 - 0.0171 \ii$ \\
  $L=12$ & $1.1398 - 0.0196 \ii$ & $1.1397 - 0.0195 \ii$ \\ 
  $L=16$ & $1.1389 - 0.0201 \ii$ & $1.1388 - 0.0203 \ii$ \\ 
  $L=20$ & $1.1388 - 0.0209 \ii$ & $1.1385 - 0.0205 \ii$ \\ 
  $L=24$ & $1.1391 - 0.0208 \ii$ & $1.1384 - 0.0206 \ii$ \\ \hline
  $L=24$ ($\hat{\rho}_{\rm A}^{\rm R}$) & $1.1635$ & $1.1633$\\
  $L=24$ ($\hat{\rho}_{\rm A}^{\rm SVD}$) & $1.1283$ & $1.1431$ \\ 
  \hline \hline
  \end{tabular}
\end{table}
\endgroup

We list the numerically obtained complex central charges $c$ in Table~\ref{tab: central-charges}.
Both the real and imaginary parts of $c$ approach their CFT values in Eq.~(\ref{eq: c-CFT}) as the system size $L$ increases, especially for $D=400$.
We get the closest imaginary part for $L=20$ and $D=300$, which seems to be a coincidence due to the competition between the finite system size effect and finite bond dimension effect;
while the former decreases the imaginary part of $c$, the latter increases it.
Notably, the critical parameters in Eq.~(\ref{eq: lambda-c}) were numerically determined in Ref.~\cite{Tang-24} up to certain uncertainty [seemingly, $\Delta(\operatorname{Re}\lambda_{\rm c})\approx 0.001,\,\Delta(\operatorname{Im}\lambda_{\rm c})\approx 0.001$] and thus our numerical results should also contain comparable uncertainty.
Considering such uncertainty, our numerically obtained complex central charges $c$ are in good agreement with theory, validating the underlying complex CFT.
Moreover, they align more closely with the CFT value in Eq.~(\ref{eq: c-CFT}) than the previous numerical result in Eq.~(\ref{eq: c-Tang-24})~\cite{Tang-24} derived from the finite-size scaling of complex eigenenergy.
Extracting the central charge from the eigenenergy scaling necessitates additional numerical fitting of the energy scale, which increases the numerical uncertainty.
By contrast, entanglement entropy, being a dimensionless quantity, does not involve such additional fitting, enhancing the precision of the numerical analysis.

\begin{figure}[tbp]
    \centering
    \includegraphics[width=0.8\linewidth]{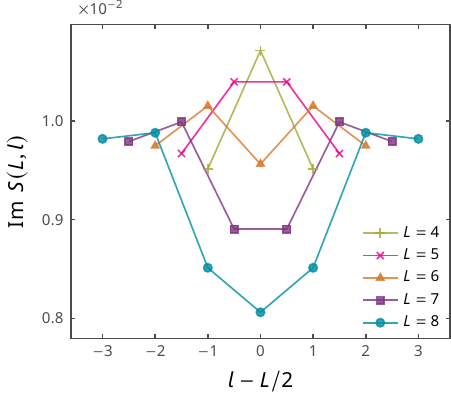}
    \caption{Imaginary part of complex entanglement entropy as a function of the subsystem size $l$, numerically obtained through the density matrix renormalization group with the bond dimension $D=400$ for the system sizes $L=4,5,6,7,8$ under the periodic boundary conditions.}
    \label{fig: complex-entropy-pbc-im-small}
\end{figure}

As shown in Fig.~\ref{fig: complex-entropy-pbc}\,(b), while the imaginary part of complex entanglement entropy is consistent with the field-theoretical description across a broad range of subsystem sizes $l$, a deviation occurs for $l \approx 1$ and $l \approx L-1$. This behavior does not seem to reflect the bulk properties of the system governed by complex CFT. As the total system size $L$ decreases, the bulk contribution diminishes, leading to the more pronounced deviation,
as illustrated in Fig.~\ref{fig: complex-entropy-pbc-im-small}.
For example, for $L=4$, the small-$l$ or large-$l$ effect
pervades the entire system, yielding a central charge with a positive imaginary part (see Table~\ref{tab: central-charges}).
For larger $L$, this deviation becomes increasingly negligible and converges toward the CFT description.

\subsection{Open boundary conditions}

\begin{figure}[b]
\centering
\includegraphics[width=\linewidth]{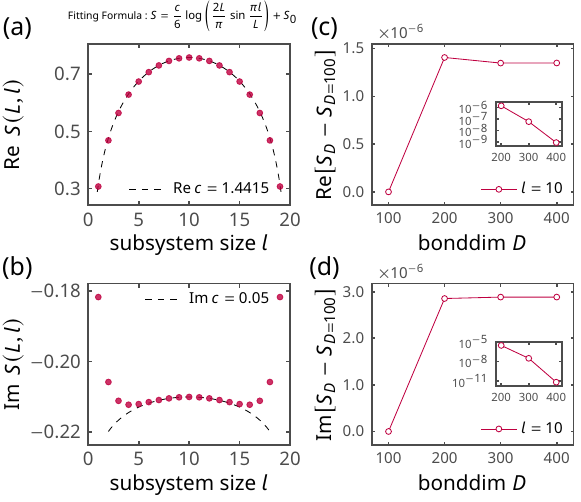}
\caption{(a)~Real and (b) imaginary parts of complex entanglement entropy as functions of the subsystem size $l$, numerically obtained through the density matrix renormalization group for the system size $L=20$ and bond dimension $D=400$ under the open boundary conditions.
We fit the numerical data using the three points in the center.
(c)~Real and (d) imaginary parts of complex entanglement entropy compared to those of $D=100$ [i.e., $S_{D} \left( L, L/2 \right) -S_{D=100} \left( L, L/2 \right)$] with respect to the bond dimension $D$.
Insets: The changes in entropy compared to the $100$ smaller bond dimension [i.e., $S_D \left( L, L/2 \right) - S_{D-100}\left( L, L/2 \right)$] for each $D = 200, 300, 400$.}
    \label{fig: complex-entropy-obc}
\end{figure}

Furthermore, we numerically obtain the complex entanglement entropy for the critical non-Hermitian five-state Potts model under the open boundary conditions.
Similar to the periodic boundary conditions, the real part of the complex entanglement entropy agrees well with the finite-size-scaling behavior of CFT~\cite{Calabrese-Cardy-04, Calabrese-Cardy-review} [Fig.~\ref{fig: complex-entropy-obc}\,(a)], 
\begin{equation}
    S^{\mathrm{OBC}} (L, l) = \frac{c}{6} \log \left( \frac{2L}{\pi} \sin \frac{\pi l}{L} \right) + S_0.
\end{equation}
However, the numerically obtained central charge yields $\mathrm{Re}\,c \approx 1.4415$, which is significantly different from that for complex CFT and from that numerically obtained under the periodic boundary conditions (see Table~\ref{tab: central-charges}). 
Additionally, the behavior of the imaginary part of the complex entanglement entropy fails to obey the description of complex CFT, as shown in Fig.~\ref{fig: complex-entropy-obc}\,(b).

The nonconformity with the CFT description is not attributed to the ill convergence in DMRG.
To demonstrate this, we show the change in complex entanglement entropy with respect to the bond dimension $D$ in Figs.~\ref{fig: complex-entropy-obc}\,(c) and \ref{fig: complex-entropy-obc}\,(d).
These numerical results suggest that increasing the bond dimension $D$ beyond $400$ should not cause a change larger than $10^{-6}$ in the entanglement entropy and thus in the obtained central charge.
While the scale of $\mathrm{Im} \left[ S_D - S_{D = 100} \right]$ for the open boundary conditions in Figs.~\ref{fig: complex-entropy-obc}\,(c) and \ref{fig: complex-entropy-obc}\,(d) is $10^{-6}$, the counterpart for the periodic boundary conditions in Figs.~\ref{fig: complex-entropy-pbc}\,(c) and \ref{fig: complex-entropy-pbc}\,(d) is $10^{-3}$.
This comparison also suggests that the ground state with open boundaries can be more effectively obtained by DMRG than that with periodic boundaries, as is normally the case with Hermitian systems.

The observed deviation from complex CFT under the open boundary conditions implies the subtle boundary effects inherent in non-Hermitian quantum many-body systems.
Notably, the expected CFT scaling behavior cannot be also obtained in a different class of nonunitary CFT with a negative central charge $c=-2$~\cite{Chang-20}. 
It is also worth noting that non-Hermitian systems can generally exhibit the extreme sensitivity to boundary conditions, a phenomenon known as the non-Hermitian skin effect~\cite{Lee-16, YW-18-SSH, Kunst-18, Yokomizo-19, Zhang-20, OKSS-20}.

\subsection{Other definitions of entanglement entropy}
    \label{subsec: other}

For comparison, we also calculate other measures of entanglement entropy for non-Hermitian systems.
In contrast to the non-Hermitian reduced density matrices $\hat{\rho}^{\RL}_{\rm A}$ in Eq.~(\ref{eq: pseudo reduced density matrix}), we choose the reduced density matrices as the Hermitian one constructed solely from the right ground state $\ket{\psi}$ [i.e., Eq.~(\ref{eq: reduced density matrix})],
\begin{equation}
    \hat{\rho}^{\rm R}_{\rm A} \coloneqq \tr_{\rm B}\,\ket{\psi} \bra{\psi},
\end{equation}
and that from SVD~\cite{parzygnat2023svdentanglemententropy},
\begin{equation}
    \hat{\rho}_{\rm A}^{\rm SVD} \coloneqq \dfrac{\sqrt{(\hat{\rho}^{\RL}_{\rm A})^\dagger \hat{\rho}^{\RL}_{\rm A}}}{\tr \sqrt{(\hat{\rho}^{\RL}_{\rm A})^\dagger \hat{\rho}^{\RL}_{\rm A}}}. \label{eq: rho-svd}
\end{equation}
Owing to Hermiticity of $\hat{\rho}^{\rm R}_{\rm A}$ and $\hat{\rho}_{\rm A}^{\rm SVD}$, the corresponding entanglement entropy is always real.
We also provide the numerically obtained real central charges in Table~\ref{tab: central-charges}.
They are less close to the real part of the CFT value than the complex central charges obtained from the non-Hermitian reduced density matrices $\hat{\rho}^{\RL}_{\rm A}$ in Eq.~(\ref{eq: pseudo reduced density matrix}), supporting the validity of complex entanglement entropy.

\begin{figure}[tbp]
    \centering
    \includegraphics[width=\linewidth]{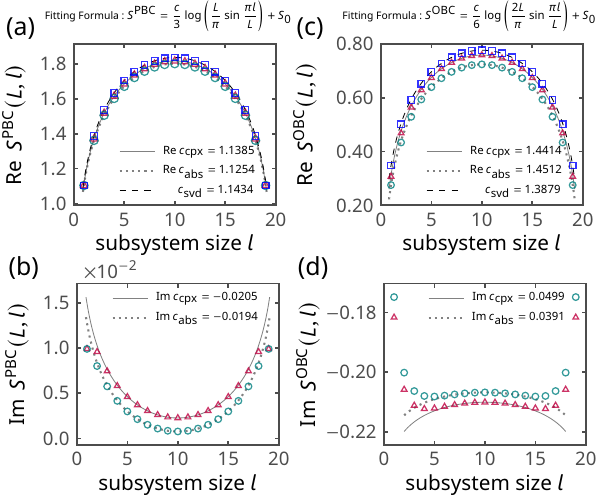}
    \caption{(a), (c)~Real and (b), (d)~imaginary parts of complex entanglement entropy $S_{\rm cpx} \in \mathbb{C}$ ($\trianglemarker$) in Eq.~(\ref{eq: pseudo entropy}), entropy $S_{\rm abs} \in \mathbb{C}$ in Eq.~(\ref{eq: abs-entropy}) ($\circlemarker$), and singular-value decomposition entropy $S_{\rm svd} \in \mathbb{R}$ ($\boxmarker$) defined from Eq.~(\ref{eq: rho-svd}) for both (a), (b)~periodic and (c), (d)~open boundary conditions. 
    The numerical results are obtained through the density matrix renormalization group with $L=20$ and $D=400$. 
    The legends show the corresponding central charges $c_{\rm cpx} \in \mathbb{C}$, $c_{\rm abs} \in \mathbb{C}$, and $c_{\rm svd} \in \mathbb{R}$ obtained by fitting the three points in the center.}
    \label{fig: compare-entropies}
\end{figure}

Additionally, in Fig.~\ref{fig: compare-entropies}, we compare 
the different definitions of entanglement entropy as a function of the subsystem size.
Specifically, we present the complex entanglement entropy in Eq.~(\ref{eq: pseudo entropy}), entropy in Eq.~(\ref{eq: abs-entropy}), and SVD entropy defined from Eq.~(\ref{eq: rho-svd}), each of which is denoted by $S_{\rm cpx} \in \mathbb{C}$, $S_{\rm abs} \in \mathbb{C}$, and $S_{\rm svd} \in \mathbb{R}$, respectively.
For the periodic boundary conditions, the real parts of $S_\mathrm{cpx}$, $S_\mathrm{abs}$, and $S_\mathrm{svd}$ coincide at $l=1,19$ with differences of the order of $10^{-4}$, but exhibit deviations of the order of $10^{-2}$ for other subsystem sizes $l$. 
Similarly, the imaginary parts of $S_\mathrm{cpx}$ and $S_\mathrm{abs}$ differ by only the order of $10^{-7}$ at $l=1,19$, while the discrepancy grows to the order of $10^{-3}$ at the center of the system. 
These observations suggest that $S_\mathrm{abs}$ and $S_\mathrm{svd}$ deviate from $S_\mathrm{cpx}$ in the regions where the features of complex CFT are prominent. 
On the other hand, for the open boundary conditions, the real parts of $S_\mathrm{cpx}$, $S_\mathrm{abs}$, and $S_\mathrm{svd}$ consistently deviate from each other with the order of $10^{-2}$ for every $l$, and the imaginary parts of $S_\mathrm{cpx}$ and $S_\mathrm{abs}$ differ by the order of $10^{-3}$ for every $l$, leaving the interpretation of these behaviors unclear.

\subsection{Comparison with $c=-2$ conformal field theory}
    \label{subsec: Petermann}

\begin{figure}[t]
\centering
\includegraphics[width=\linewidth]{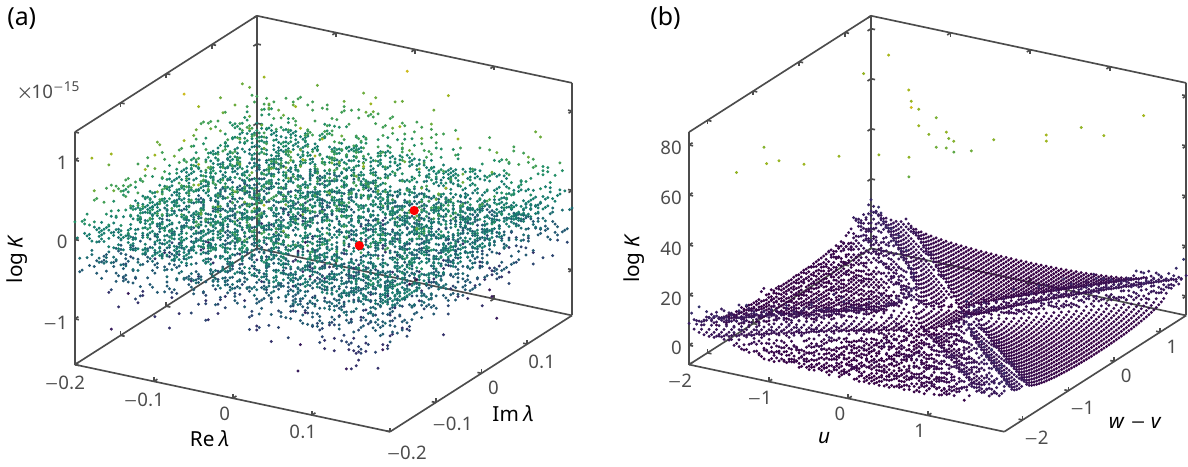}
\caption{(a)~Petermann factor $K$ for the ground state of the non-Hermitian five-state Potts model ($L=2$, $J=1$, $h=1$). 
The two large red points show the Petermann factors at the critical points $\lambda = 0.079 \pm 0.060\ii$. 
(b)~Petermann factor $K$ for the ground state of the non-Hermitian Su-Schrieffer-Heeger model in Eq.~\eqref{eq: nh-ssh} ($L=8, v=0.4$, periodic boundary conditions, half filling). 
The divergence occurs along the lines $v \pm w=\pm u$.
The points at which the Petermann factor diverges are excluded from the plot.}
    \label{fig: Petermann}
\end{figure}

The complex CFT studied in this work exhibits fundamentally distinct finite-size-scaling properties from another class of nonunitary CFT characterized by negative central charges.
The critical non-Hermitian Potts model retains an energy gap between the ground state and first excited state for finite system sizes, enabling reliable extraction of the complex central charge and associated scaling dimensions~\cite{Tang-24}.
In this work, we demonstrate that a complex-valued generalization of entanglement entropy also captures this CFT scaling behavior.
In contrast, identification of $c < 0$ via the finite-size scaling of eigenenergy is elusive, unlike unitary CFT with $c > 0$.
Instead, the $c = -2$ CFT scaling behavior lies in the presence of exceptional points that emerge even in finite-size systems~\cite{Chang-20}. 

To further substantiate the distinct finite-size-scaling behavior, we numerically calculate the Petermann factor~\cite{Petermann-79}
\begin{equation}
    K \coloneqq \dfrac{\braket{\psi|\psi} \langle\!\braket{\psi|\psi}\!\rangle}{\lvert \langle\!\braket{\psi|\psi} \rvert^2}.
\end{equation}
The Petermann factor quantifies the nonorthogonality between right and left eigenstates, significantly increasing near exceptional points and diverging exactly at such points~\cite{Wiersig-23}.
In addition to the non-Hermitian Potts model, we also investigate a non-Hermitian extension of the Su-Schrieffer-Heeger (SSH) model~\cite{Chang-20},
\begin{equation}    \label{eq: nh-ssh}
\begin{split}
    \hat{H} &= \sum_{k} \hat{c}_k^{\dag} H \left( k \right) \hat{c}_k, \\
    H \left( k \right) &= \begin{pmatrix}
        \ii u  & v + w e^{-\ii k} \\
        v + w e^{\ii k} & -\ii u
    \end{pmatrix}, 
\end{split}
\end{equation}
where $u, v, w$ are real-valued parameters, and $\hat{c}_k$ ($\hat{c}_k^{\dag}$) annihilates (creates) a fermion with momentum $k$.
This model exhibits the $c=-2$ CFT behavior at the critical points, as revealed by the finite-size scaling of entanglement entropy.
In Fig.~\ref{fig: Petermann}, we present the numerically computed Petermann factors $K$ for both non-Hermitian Potts and SSH models around their critical points.
While $K$ remains almost one around the critical point in the finite-size non-Hermitian Potts model, it exhibits a pronounced enhancement as the system approaches the critical points in the non-Hermitian SSH model, further underscoring the distinct finite-size-scaling behavior.

\section{Discussion}
    \label{sec: discussion}

In this work, we demonstrate that complex entanglement entropy effectively characterizes complex CFT.
We introduce non-Hermitian reduced density matrices derived from the combination of right and left eigenstates.
As a prototypical model described by complex CFT, we investigate the five-state Potts model with a non-Hermitian perturbation.
Using DMRG, we numerically obtain the right and left ground states at a critical point and calculate complex entanglement entropy.
We confirm the consistency with the CFT scaling behavior associated with a complex central charge.

While complex entanglement entropy captures complex CFT, it does not retain certain key properties of conventional entanglement entropy, such as strong subadditivity.
Further research is needed to put a clear interpretation on complex entanglement entropy.
In Hermitian systems, the entanglement spectrum~\cite{Ryu-06, Li-Haldane-08}, beyond the entanglement entropy, often encodes significant information about the underlying unitary field theory~\cite{Lauchli-13, Cho-17}.
Similarly, the spectrum of the non-Hermitian reduced density matrix in Eq.~(\ref{eq: pseudo reduced density matrix}) can reveal insights into the underlying nonunitary field theory.
Moreover, the complex entanglement entropy explored in this work is recognized as pseudo-entropy in recent studies of high energy physics~\cite{Nakata-21, Mollabashi-21, *Mollabashi-21R, Murciano-22, Guo-23, Doi-23, *Doi-23JHEP, Narayan-23, Kanda-23, Kanda-24}.
It is worthwhile to study the implications of our results within this context, such as the relevance to the de Sitter (dS)/CFT correspondence.
Additionally, we note that non-Hermitian density matrices are utilized in generic tensor network formulations (see, for example, Ref.~\cite{Tang-23} and references therein).

It warrants further investigation to construct other non-Hermitian models featuring complex fixed points and study their nonunitary critical phenomena.
For example, the Kondo problem with a complexified coupling parameter was perturbatively shown to exhibit the reversible renormalization-group flow,
potentially implying complex-valued scaling dimensions and an underlying complex CFT~\cite{Nakagawa-18, Han-23}.
Non-Hermiticity of the Kondo interaction is physically rooted in inelastic collisions and the resultant losses of ultracold atoms.
More broadly, non-Hermiticity arises from the exchange of energy and particles with the external environment~\cite{Konotop-review, Christodoulides-review}.
Therefore, a deeper understanding of complex CFT should also provide insights into phase transitions and critical phenomena in open quantum systems.

\begin{acknowledgments}
We thank Yoshiki Fukusumi, Kansei Inamura, Taishi Kawamoto, Tokiro Numasawa, Masaki Oshikawa, Shinsei Ryu, Yin Tang, Kenya Tasuki, Kazuki Yamamoto, and Mengyang Zhang for helpful discussions.
We acknowledge the long-term workshop ``Recent Developments and Challenges in Topological Phases'' (YITP-T-24-03) held at Yukawa Institute for Theoretical Physics (YITP), Kyoto University.
K.K. is supported by MEXT KAKENHI Grant-in-Aid for Transformative Research Areas A ``Extreme Universe" No.~JP24H00945.
A part of the computation in this work was done using the facilities of the Supercomputer Center, the Institute for Solid State Physics, the University of Tokyo.
\end{acknowledgments}

\bibliography{ref.bib}

\end{document}